\documentclass[pra,aps,twocolumn,10pt,showpacs,groupedaddress,superscriptaddress,floatfix,notitlepage]{revtex4-2}
\usepackage{amsmath,amsfonts,amssymb,graphics,graphicx,epsfig,color,times}
\usepackage[utf8x]{inputenc}
\usepackage{color}
\usepackage{bbm, dsfont} 
\usepackage{subfigure}
\usepackage{hyperref}
\usepackage{mathrsfs}
\usepackage{verbatim}
\begin{document}
\title{Quantum correlations of localized atomic excitations in a disordered atomic chain}
\author{H. H. Jen}
\email{sappyjen@gmail.com}
\affiliation{Institute of Atomic and Molecular Sciences, Academia Sinica, Taipei 10617, Taiwan}

\date{\today}
\renewcommand{\r}{\mathbf{r}}
\newcommand{\f}{\mathbf{f}}
\renewcommand{\k}{\mathbf{k}}
\def\p{\mathbf{p}}
\def\q{\mathbf{q}}
\def\bea{\begin{eqnarray}}
\def\eea{\end{eqnarray}}
\def\ba{\begin{array}}
\def\ea{\end{array}}
\def\bdm{\begin{displaymath}}
\def\edm{\end{displaymath}}
\def\red{\color{red}}
\pacs{}
\begin{abstract}
Atom-waveguide interface mediates significant and long-range light-matter interactions through the guided modes. In this one-dimensional system, we theoretically investigate the excitation localization of multiple atomic excitations under strong position disorders. Deep in the localization side, we obtain the time evolutions of quantum correlations via Kubo cumulant expansions, which arise initially and become finite and leveled afterward, overtaking the ones without disorders. This indicates two distinct regimes in time: before the onset of excitation localization, the disorders engage the disturbance of quantum correlations, which is followed by disorder-assisted build-up of quantum correlations that maintain at a later stage owing to the absence of excitations diffusion. The crossing of distinct regimes is pushed further in time for longer-range correlations, which indicates a characteristic timescale needed for disorders to sustain them. We also explore the effect of directionality of couplings and resonant dipole-dipole interactions, which can drive the system toward the delocalized side when it is under chiral couplings or large dipole-dipole interaction strengths. The time-evolved quantum correlations can give insights to the studies of few-body localization phenomenon and nonequilibrium dynamics in open quantum systems.                 
\end{abstract}
\maketitle
\section{Introduction}

Anderson transition \cite{Anderson1958, Evers2008} connects two distinct phases from a metal to an insulator in an electronic system under random potentials. In an analogy to Anderson transition, many closed and interacting quantum systems share this similar delocalization to localization transition, for example, in an electron gas with attractive interactions \cite{Giamarchi1988}, bosons under periodic potentials \cite{Fisher1989}, and Bose-Einstein condensates \cite{Clement2005, Fort2005}. Aside from these representative many-body systems, photon localization can also emerge in a disordered media \cite{Wiersma2013} or under resonant dipole-dipole interactions \cite{Akkermans2008}. Two common features underlying these localization phenomena are the interference mechanism from multiple scattering of quantum particles and the absence of their transport \cite{Anderson1958} or conductivity \cite{Nandkishore2015}.  

Recently, a new dynamical phase of many-body localization emerges \cite{Bardarson2012, Agarwal2015, Schreiber2015, Vosk2015, Nandkishore2015, Choi2016, Roushan2017, Abanin2019}, which in the presence of atom-atom interactions shows distinct nonergodic behavior in highly excited eigenstates with a logarithmic spread of entanglement entropy in time \cite{Bardarson2012, Schreiber2015}. Though ultimately unbounded as shown in the numerical study of number entropy \cite{Kiefer2020}, the many-body localized phase within experimentally accessible time of interest still presents rich nonequilibrium dynamics and unique stationary phase diagram with an interplay between interaction and disorder strengths \cite{Schreiber2015, Choi2016}. Furthermore, a different universality class of non-Hermitian many-body localization \cite{Hamazaki2019} can be identified by complex-to-real transition of the eigenvalues in a model of interacting bosons with asymmetric hopping. This extends the localization phenomenon to the realm of open quantum systems, and it can be associated with new signatures of level-spacing statistics from complex eigenspectrum \cite{Hamazaki2019, Hamazaki2020, Sa2020} that are closely related to Hermitian random matrix theory \cite{Haake2010}.

Since an open quantum system in general involves dissipation and decoherence \cite{Hammerer2010}, the localization of quantum particles fades away as time evolves, as well as their properties of entanglement entropy. Therefore, it is challenging to unravel clear signatures of localization under a dissipative environment. In a setup of light-matter interacting quantum interface with collective dipole-dipole interactions \cite{Kien2005, Solano2017}, a strongly-coupled atom-waveguide system \cite{Chang2018, Corzo2019} presents subradiant dynamics \cite{Albrecht2019, Jen2020_subradiance, Kumlin2020} which sustains for a longer lifetime and thus suffices to simulate Anderson-like localization of a quenched single atomic excitation \cite{Jen2020_disorder, Jen2021_crossover}. The atom-waveguide system has been shown to manifest mesoscopic entanglement \cite{Tudela2013}, photon-photon correlations \cite{Mahmoodian2018, Jeannic2021}, long-range correlated spin dimers \cite{Ramos2014, Pichler2015}, and bounded multiatom excitations \cite{Jen2021_bound}. It also allows tunable nonreciprocal couplings in the guided modes of the waveguide to realize a new research paradigm of chiral quantum optics \cite{Mitsch2014, Lodahl2017, Jen2020_PRR} and topological waveguide quantum electrodynamics \cite{Kim2021, Sheremet2021}

Here we theoretically study quantum correlations via Kubo cumulant expansions \cite{Kubo1962} in an atomic chain coupled to the waveguide. When the system is subject to position disorders, we numerically obtain these time-evolved second- and third-order nonclassical correlations as measures to investigate the localization of quenched two and three excitations, respectively. We find that a finite and leveled quantum correlation emerges under strong disorders and overtakes the one without. This results in two distinct regimes of time, where initially the disorders reduce the degree of quantum correlations and maintain them later on due to the absence of excitations diffusion. We further explore longer-range correlations that can be sustained in our system and the effect of chiral couplings and light-induced dipole-dipole interactions. We find that these effects can drive the system toward the delocalization side when it is under chiral couplings with finite directionality or large dipole-dipole interaction strengths. The time-evolved quantum correlations can provide an alternative and additional probe of the localization phenomenon in open quantum systems.

In this article, we focus on a system of disordered atomic chain in an atom-waveguide interface as shown in Fig. \ref{fig1}. We introduce the theoretical model of the interface under disorders in Sec.$~$II, and in Sec.$~$III, we present the localized two and three atomic excitations under strong disorders. In Sec.$~$IV, we define the average density-density and third-order correlations to analyze multiple excitations localization in time. Finally we discuss our results and conclude in Sec.$~$V.    

\section{Theoretical model of an atom-waveguide interface under disorders}

\begin{figure}[t]
\centering
\includegraphics[width=8.5cm,height=6cm]{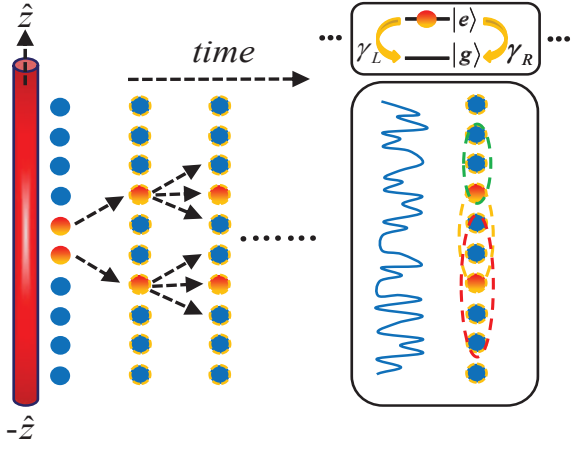}
\caption{Schematic plot for localization of two atomic excitations. The atom-waveguide coupled system consists of multiple two-level quantum emitters which mediate long-range dipole-dipole interactions via the guided modes (evanescent waves) in the waveguide. The initially excited atoms decay from the excited ($|e\rangle$) to the ground states ($|g\rangle$) with nonreciprocal decay channels ($\gamma_{L}\neq\gamma_{R}$) to the left ($-\hat z$) or right side ($\hat z$) of the atomic chain. These spin excitations under dissipations and spin-exchange interactions transport and spread to the edges of the chain, but become localized once position disorders are introduced (denoted by a blue solid line in the lower right inset plot), along with significant and sustaining long-range quantum correlations (denoted by dashed lines in colors).}\label{fig1}
\end{figure}

To investigate the quantum correlations in a disordered atomic chain, we focus on an atom-waveguide system with nonreciprocal decay rates ($\gamma_L\neq \gamma_R$), which can be effectively formed by an atomic chain of two-level atoms coupling with the guided modes. In Fig. \ref{fig1}, a schematic plot of this interface is shown, and the system dynamics can be described by a density matrix $\rho$ in Lindblad forms ($\hbar$ $=$ $1$) \cite{Pichler2015, Jen2020_disorder, Jen2021_bound}, 
\bea
\frac{d \rho}{dt}=-i[H_L+H_R,\rho]+\mathcal{L}_L[ \rho]+\mathcal{L}_R[\rho],\label{rho}
\eea
where the coherent and Hermitian part is
\bea
H_{L(R)} =&& -i\frac{\gamma_{L(R)}}{2} \sum_{\mu<(>)\nu}^N\left(e^{ik_s|r_\mu-r_\nu|} \sigma_\mu^\dag\sigma_\nu-\textrm{H.c.}\right),\label{LR}
\eea
and the dissipative part in Lindblad forms is
\bea
\mathcal{L}_{L(R)}[\rho]=&&-\frac{\gamma_{L(R)}}{2} \sum_{\mu,\nu}^N e^{\mp ik_s(r_\mu-r_\nu)} \left(\sigma_\mu^\dag \sigma_\nu \rho + \rho \sigma_\mu^\dag\sigma_\nu \right.\nonumber\\
&&\left.-2\sigma_\nu \rho\sigma_\mu^\dag\right).
\eea

The dipole operators are $\sigma_\mu^\dag$ $\equiv$ $|e\rangle_\mu\langle g|$ with $|g\rangle$ and $|e\rangle$ for the ground and excited states, which can be interpreted as an effective spin $1/2$ system. The $k_s$ denotes the wave vector in the guided mode, and we use $\gamma_{L(R)}$ to characterize the directional coupling rates to the left ($L$) or right ($R$), from which $D\equiv(\gamma_R-\gamma_L)/(\gamma_R+\gamma_L)$ denotes a normalized quantity of directionality \cite{Mitsch2014}. The effective model of Eq. (\ref{rho}) can be obtained with Born-Markov approximation by treating the reservoirs only in the allowed dimension determined by the waveguide \cite{Tudela2013}. The intrinsic timescale of the system can be identified as $\gamma_R$ $+$ $\gamma_L$ $=$ $\gamma$ $\equiv$ $2|dq(\omega)/d\omega|_{\omega=\omega_{eg}}g_{k_s}^2L$, a total decay rate, where $|dq(\omega)/d\omega|$ denotes the inverse of group velocity with a resonant wave vector $q(\omega)$, $g_{k_s}$ the coupling constant, and $L$ the quantization length. 

In this atomic chain with equal interparticle distances, we define $\xi$ $\equiv$ $k_s |r_{\mu+1}-r_{\mu}|$ to specify the strength of the one-dimensional resonant dipole-dipole interactions, which is responsible of superradiant emissions at a long distance \cite{Solano2017} and leads to subradiant dynamics \cite{Jen2020_subradiance} or bound multimers \cite{Jen2021_bound} when $\xi$ $\sim$ $\pi$. The system dynamics can be fully determined by solving Eq.$~$(\ref{rho}), where under a driven condition the many-body dark states and nonlocal multimers can be created \cite{Pichler2015}, and a steady-state phase diagram of bi-edge or -hole excitations can be identified as well \cite{Jen2020_PRR}.  

When multiple atoms are quenched to the excited states initially, we need a complete Hilbert space of $C^N_M$ bare states of $M$ atomic excitations in $N$ atoms to simulate the system dynamics, where $C$ denotes the binomial coefficient. The bare states can be expressed by $|\phi_q\rangle$ $=$ $\sigma_j^\dag \sigma_{k>j}^\dag ...\sigma_{l}^\dag\sigma_{m>l}^\dag|0\rangle$ as the labeled $q$th state basis for $M$ excitations in general. Under this bare state basis, we can numerically calculate $\rho(t)=|\Psi(t)\rangle\langle\Psi(t)|$ from Eq. (\ref{rho}), where $|\Psi(t)\rangle$ $=$ $\sum_{q=1}^{C^N_M}a_q(t)|\phi_q\rangle$ and the corresponding probability amplitudes $a_q(t)$ can be solved from \cite{Jen2017_MP, Jen2021_bound}
\bea
\dot{a}_q(t)=\sum_{s=1}^{C^N_M}V_{q,s} a_s(t). \label{a}
\eea   
The diagonal elements of the coupling matrix $V_{q,s}$ are $-M\gamma/2$ directly from $M$ excited atoms, and its off-diagonal parts can be obtained from the interaction kernel $\bar V$ under single-excitation bare states,  
\bea
\bar V_{\mu,\nu}=\left\{\begin{array}{lr}
    -\gamma_Le^{-ik_s|r_{\mu}-r_\nu|},~\mu<\nu\\
		-\frac{\gamma}{2}\delta_{\mu,\nu}\\
		-\gamma_Re^{-ik_s|r_{\mu}-r_\nu|},~\mu>\nu
\end{array}\right..\label{V}
\eea
$V_{q,s}$ is finite and can be assigned by $\bar V_{\mu,\nu}$ as long as only the $\mu$th excited atom in the multiply-excited bare states $|\phi_{q}\rangle$ does not overlap with the excited $\nu$th one in $|\phi_{s}\rangle$. This reflects the feature of pairwise couplings in dipole-dipole interactions, and $V_{q,s}$ becomes null when more than two excited atoms in $|\phi_{q}\rangle$ do not coincide with the ones in $|\phi_{s}\rangle$. For example, between $|e_1, e_2, g_3, g_4\rangle$ and $|g_1, g_2, e_3, e_4\rangle$ or $|e_1, e_2, e_3, g_4, g_5\rangle$ and $|g_1, g_2, e_3, e_4, e_5\rangle$, for the cases of $M=2$ with $N=4$ and $M=3$ with $N=5$, respectively, $V_{q,s}$ is vanishing. On the other hand, for example of the case of $M=2$ with $N=4$, $V_{q,s}$ is finite between $|e_1, e_2, g_3, g_4\rangle$ and $|e_1, g_2, e_3, g_4\rangle$ or $|g_1, e_2, e_3, g_4\rangle$ and $|g_1, e_2, g_3, e_4\rangle$. 

We have applied the above formalism in studying the subradiant emissions from multiply-excited atoms in free space \cite{Jen2017_MP} and shape-preserving dimers and trimers of atomic excitations in an atomic array with nonreciprocal couplings \cite{Jen2021_bound}. The essential information of excitation population can then be calculated in a straightforward way when we arrange these bare states into $(N-M+1)$ sectors \cite{Jen2017_MP}, where each sector indicates one increment of the index of the first atomic excitation when we order $N$ atomic positions as $r_1$ $<$ $r_2$ $<...<$ $r_{N-1}$ $<$ $r_N$. Therefore, in the first sector the bare states are $|e_1e_2$ $...$ $e_M g_{M+1}...g_N\rangle$, $|e_1e_2$ $...$ $e_{M-1} g_M e_{M+1} g_{M+2}$ $...g_N\rangle$, $\cdots$, $|e_1e_2$ $...$ $e_{M-1} g_M$ $...$ $g_{N-1}e_N\rangle$, until $(M-1)$ excited atoms move to the end of the chain as $|e_1g_2$ $...$ $g_{N-M+1} e_{N-M+2}$ $...$ $e_N\rangle$. The second sector starts from $|g_1e_2e_3$ $...$ $e_{M+1} g_{M+2}$ $...g_N\rangle$, the rest can be constructed sequentially, and the last sector involves only one bare state $|g_1g_2$ $...$ $g_{N-M}e_{N-M+1}$ $...e_N\rangle$. From the solutions of $|\Psi(t)\rangle$, we are able to obtain the excitation population $P_m(t)$ $=$ Tr$(\rho(t)\sigma^\dag_m\sigma_m)$, where a conservation of total spin excitations should be $\sum_{m=1}^N P_m(t)$ $=$ $M$. In a disordered atomic media we consider here, the atoms experience effectively onsite phase disorders $W_\mu$ $\in$ $W[-1,1]$ and $W\in[0,\pi]$ owing to the position fluctuations, which reflects in a deviation of $\xi$ in $H_{L(R)}$ and $\mathcal{L}_{L(R)}[\rho]$, and accordingly in Eq. (\ref{a}) with an extra $e^{-i(W_\mu-W_\nu)}$ in $V_{q,s}$ between the $\mu$th and $\nu$th atoms \cite{Jen2020_disorder}.

In the following two sections, we numerically obtain the average atomic excitation populations and their quantum correlations under positions disorders. We will see how the directionality $D$, light-induced dipole-dipole interaction strengths, and disorder strengths $W$ modify the localization behaviors from the localization side, from which we can further identify how the system can be driven toward the delocalization. Since the system is under dissipations, we always compare our results with the ones without disorder, which provides a reference of the thermalized phase.     

\section{Localization of multiple spin excitations under disorders}

To investigate the effect of position disorders on the diffusion of multiple atomic excitations, we focus on the localization side under strong disorders. Throughout this paper, we numerically calculate the excitation populations $\langle P_m(t)\rangle$ and quantum correlations in the next section under $400$ ensemble average of disorders. The $\langle P_m(t)\rangle$ has reached good enough convergence within time of interests, where the total population deviates by less than $4\%$ from the one under $800$ realizations.   

\begin{figure}[t]
\centering
\includegraphics[width=8.5cm,height=4.5cm]{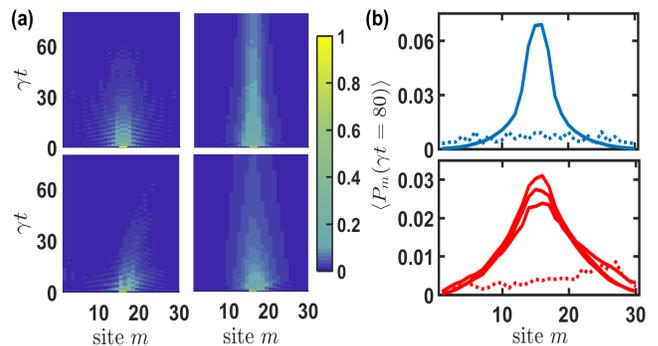}
\caption{Time evolutions of excitation population $\langle P_m(t)\rangle$ for initial two atomic excitations side by side in an atomic chain of $N=30$ at $\xi=\pi/8$. (a) The upper and lower plots represent the cases of $\langle P_m(t)\rangle$ for $D=0$ and $0.5$, respectively, with position disorders $W=0.8\pi$ (right) or without (left). (b) The excitation population $\langle P_m(t)\rangle$ at $\gamma t=80$ in the upper and lower plots respectively correspond to the cases of right plots in (a), where localized excitation populations (solid line) emerge under strong disorders comparing the delocalized ones (dotted line). Three solid curves in the lower plot show that a localization of atomic excitations becomes more significant as $W/\pi$ increases from $0.1$, $0.5$, to $0.8$.}\label{fig2}
\end{figure}

In Fig. \ref{fig2}, we show an example of the localization in an atomic chain with initial two spin excitations side by side at the center. For a finite nonreciprocal coupling ($D\neq 0$), the spin excitations tend to transport to the end of the chain under no disorders, whereas in the reciprocal case ($D= 0$) the excitation populations dissipate symmetrically. Both cases without disorders show interference fringes owing to the finite $\gamma_{L(R)}$, which can be attributed to multiple spin exchanges to the left and right during the spin transport. These fringes disappear under strong disorders, which associates with significant excitation populations for long time compared to the disorder-free case. The case with reciprocal couplings indicates more localized spin excitations in their peak values and population distribution at the same disorder strength. This suggests a deeper localized phase for $D=0$, or in other words for a finite $D$, it requires a larger disorder strength to enter the excitation localization phase. A similar situation can also be found in Anderson-like localization in a chirally-coupled atomic chain \cite{Jen2020_disorder}, where a larger $D\sim 1$ ($D=1$) leaves less (no) room for the excitation localization.  

\begin{figure}[b]
\centering
\includegraphics[width=8.5cm,height=4.5cm]{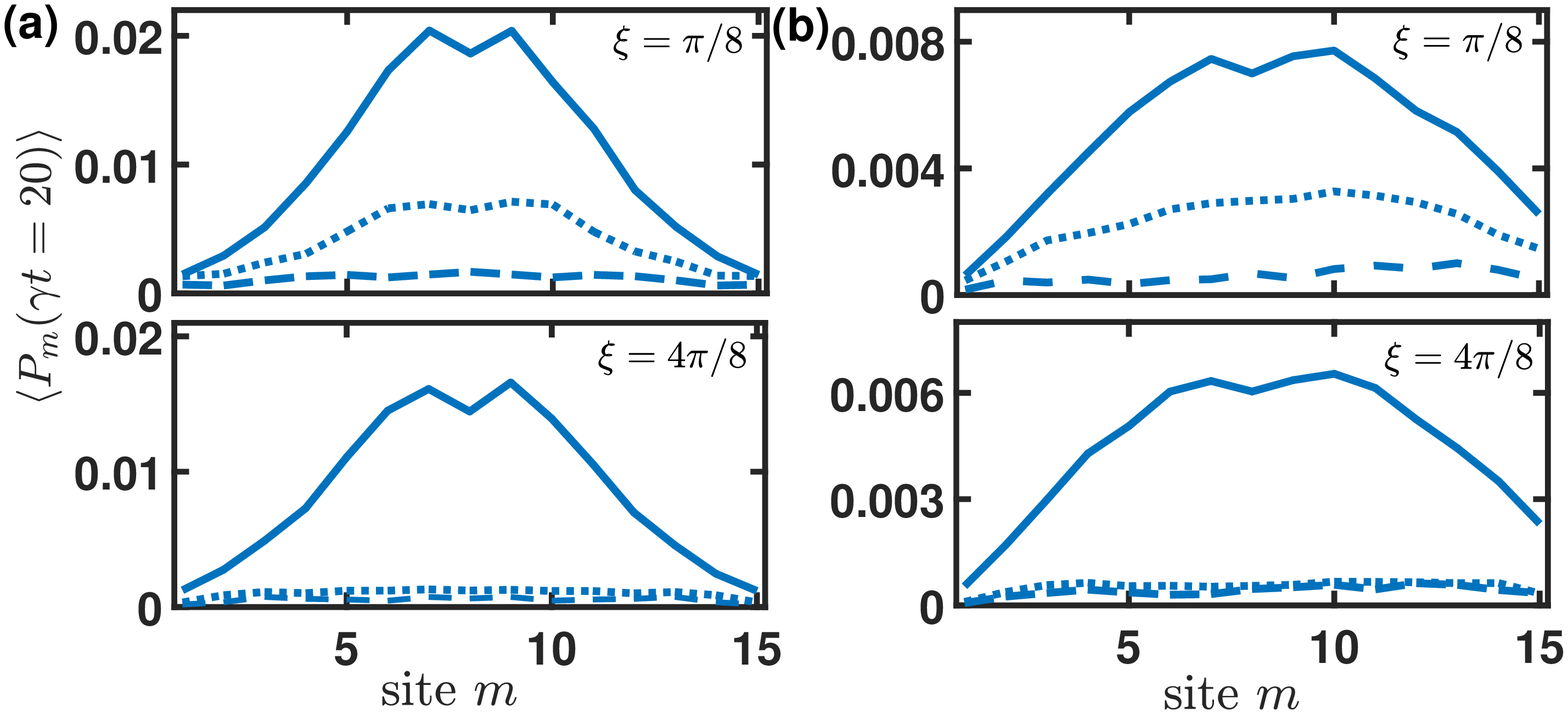}
\caption{Localization of three atomic excitations side by side. The atomic excitation populations $\langle P_m(t)\rangle$ are plotted in an atomic array with $N=15$ with different interaction strengths $\xi$ for (a) $D=0$ and (b) $0.5$, respectively, where disorder strengths are $W/\pi=0$ (dash line), $0.1$ (dotted line), and $0.8$ (solid line). From the localization side, it requires a larger disorder strength to reach significant population localization when $\xi$ or $D$ is larger.}\label{fig3}
\end{figure}

For three spin excitations under strong disorders as a comparison, we show their localization behaviors in Fig. \ref{fig3} with a larger $\xi$. Similar to Fig. \ref{fig2}, a reciprocal coupling case has a more pronounced localization of excitations along with sharper wings in their population distributions. Furthermore, as $\xi$ increases in the lower plots of Figs. \ref{fig3}(a) and \ref{fig3}(b), it takes a larger disorder strength to enter the localization phase, where the localized excitation populations are suppressed in their peak values, respectively. This indicates that the light-induced dipole-dipole interactions drive the system toward a delocalized side for low disorder strengths, which shares a similar feature for the many-body localization of two-dimensional bosons under disordered potentials \cite{Choi2016}. In contrast to the localized phase of single excitation \cite{Jen2020_disorder}, where the interactions play negligible roles for $D\gtrsim 0.5$, we find here instead that the interaction suppresses the onset of system's localization when multiple atoms are excited. This also shows a sign of gradual dominance of interactions over the directionality of nonreciprocal couplings from the localization side in a few-body localization from bottom up. 

Next, we look into the system dynamics and introduce the average quantum correlations as complementary but essential measures to investigate the localization of multiply-excited atoms in an atomic chain.  

\section{Average long-range quantum correlations}

Here we utilize the correlation functions via Kubo cumulant expansions \cite{Kubo1962} to further investigate the localization dynamics, which provide the information of quantum correlations that have no classical counterparts. The cumulant expansion provides a useful theoretical foundation that relates high-order quantum correlations with lower-order cumulants of correlations. In addition, we apply the average second-order and modified third-order correlation functions \cite{Jen2021_bound} to evaluate the bulk properties of the atomic array \cite{Keesling2019}. They are defined respectively as, 
\bea
\langle G^{(2)}(r)\rangle\equiv&&\sum_j\frac{\langle n_j n_{j+r}\rangle-\langle n_j \rangle\langle n_{j+r}\rangle}{N-r},\\
\langle G^{(3)}\rangle\equiv&&\sum_j\frac{\langle n_j n_{j+1} n_{j+2}\rangle-\langle n_j \rangle\langle n_{j+1}\rangle\langle n_{j+2}\rangle}{N-2},
\eea 
where $r$ denotes a specific distance in quantum correlations for doubly-excited spin diffusion, and they are averaged by $(N-r)$ of such correlations to represent the bulk properties. We note that there are three extra second-order cumulants in $\langle G^{(3)}\rangle$, which are $(\langle n_j n_{j+1}\rangle$ $-\langle n_j \rangle\langle n_{j+1}\rangle)$ $\langle n_{j+2}\rangle$, $(\langle n_j n_{j+2}\rangle$$~$$-\langle n_j \rangle\langle n_{j+2}\rangle)$ $\langle n_{j+1}\rangle$, and $(\langle n_{j+1} n_{j+2}\rangle$ $-\langle n_{j+1} \rangle\langle n_{j+2}\rangle)$ $\langle n_{j}\rangle$, and that is why we called it the modified third-order correlation function. This is particular useful to calculate the quantum correlations for three excited atoms initially side by side, and it can be comparable directly to experimental observations. 

\begin{figure}[t]
\centering
\includegraphics[width=8.5cm,height=4.5cm]{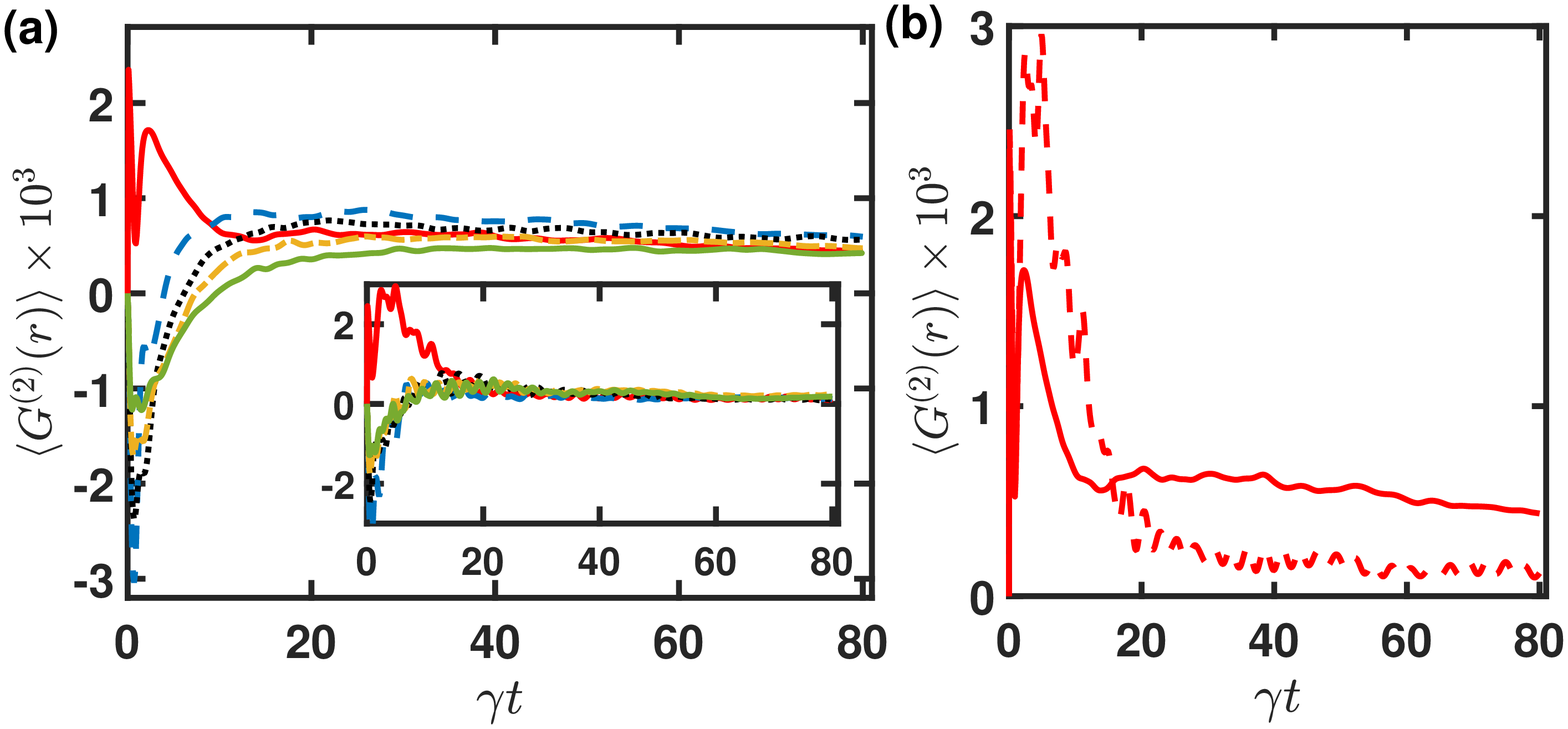}
\caption{Average second-order correlations $\langle G^{(2)}(r)\rangle$ at various separated sites $r$ deep in the localization side. Under the same system parameters in the lower right plot of Fig. \ref{fig2}(a), $\langle G^{(2)}(r)\rangle$ at $r$ $=$ $1$ (red solid line), $2$ (blue dashed line), $3$ (black dotted line), $4$ (yellow dash-dotted line), and $5$ (green solid line declining initially) become leveled as time evolves comparing the case without disorders in the inset plot correspondingly. (b) The $\langle G^{(2)}(r=1)\rangle$ under disorders (red solid line) crosses the one without (red dashed line) and stays finite as time evolves.}\label{fig4}
\end{figure}
\begin{figure}[b]
\centering
\includegraphics[width=8.5cm,height=4.5cm]{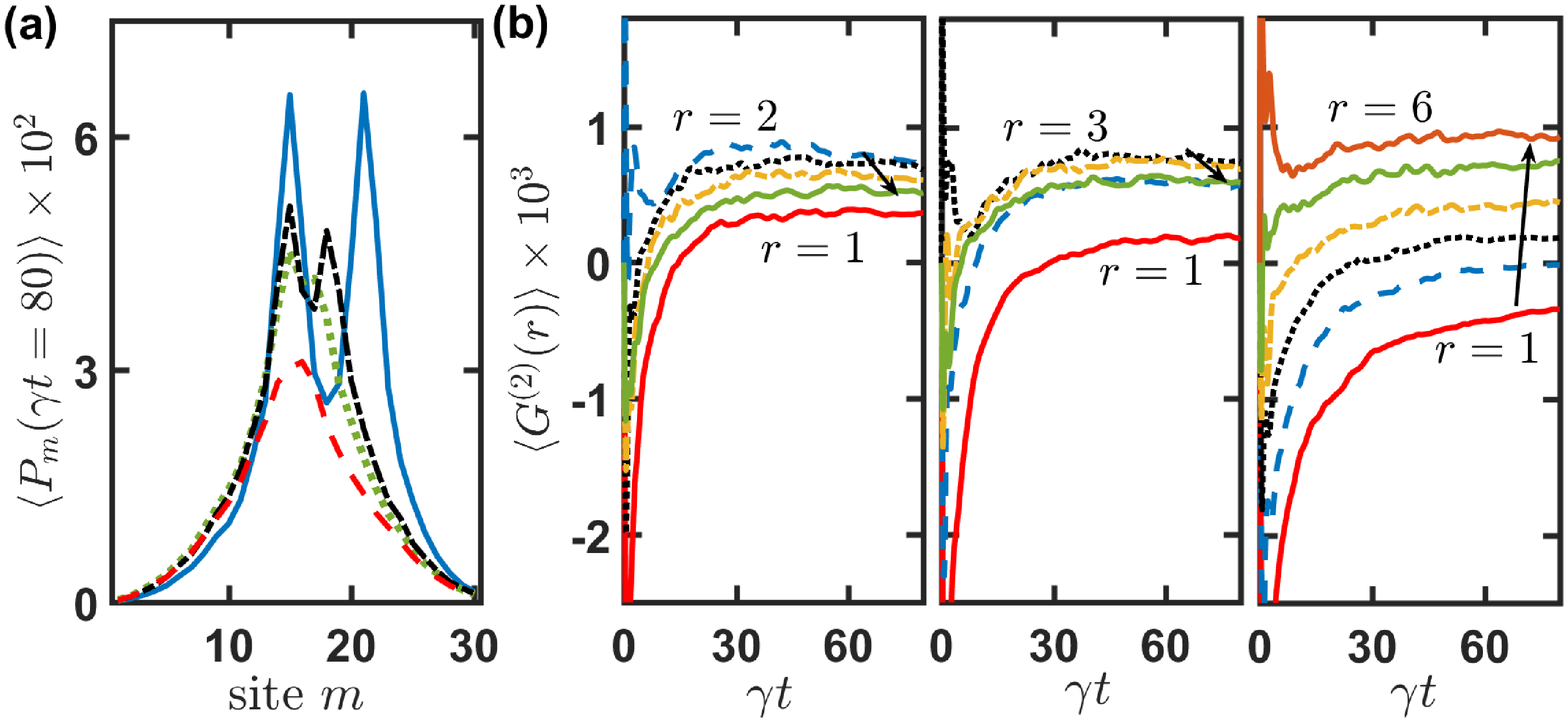}
\caption{Localized excitation populations and $\langle G^{(2)}(r)\rangle$ under various initial double atomic excitations in the localization side. (a) Localized atomic excitation populations are shown, where respective peaks are correlated to their initialized excitations with separated lattice sites of zero ($r=1$, red dashed line), one ($r=2$, green dotted line), two ($r=3$, black dash-dotted line), and five ($r=6$, blue solid line). (b) The average second-order correlations $\langle G^{(2)}(r)\rangle$ are plotted for initial double atomic excitations with separated lattice sites of one ($r=2$, left), two ($r=3$, middle), and five ($r=6$, right), respectively, that is $\sigma_j^\dag\sigma_{j+2}^\dag|0\rangle$, $\sigma_j^\dag\sigma_{j+3}^\dag|0\rangle$, and $\sigma_j^\dag\sigma_{j+6}^\dag|0\rangle$. The arrows show the $\langle G^{(2)}(r)\rangle$ with an increasing $r$, where the line styles are the same as in Fig. \ref{fig4}(a), and other system parameters are the same as in Fig. \ref{fig4}.}\label{fig5}
\end{figure}

In Fig. \ref{fig4}, we first show the results of second-order correlations for the case of $D=0.5$ in Fig. \ref{fig2}(a). At $\gamma t\gtrsim 20$, the correlations decay and almost vanish under no disorders, while $\langle G^{(2)}(r)\rangle$ sustains for long time with various separated distances $r$ under strong disorders. This provides additional information of the relation between time-evolving quantum correlations and the localization of atomic excitations. In Fig. \ref{fig4}(b), we specifically plot the quantum correlations side by side and find that there are two distinct regimes in time separated by a crossing between the correlations with and without disorders. Before the onset of excitation localization at $\gamma t\sim 20$, $\langle G^{(2)}(r=1)\rangle$ without disorders shows a more pronounced value than the one under strong disorders. We attribute it as the engaging process when disorders disturb the build-up of quantum correlations before the atomic excitations localize. This is followed by disorder-assisted preservation of quantum correlations that maintain for long time owing to the absence of excitations diffusion.

Before looking into the detail of crossing times of various quantum correlations in different correlation lengths, we next show the effect of initially separated double excitations in Fig. \ref{fig5}. The localization of initial double excitations can be seen in Fig. \ref{fig5}(a), where they preserve the information of excitation patterns in their populations. Furthermore, the corresponding quantum correlations $\langle G^{(2)}(r)\rangle$ shows a maximum highly correlated to their initial separations. For other $\langle G^{(2)}(r)\rangle$ with smaller or larger distances than the initial one, they rise up and develop as time evolves but with smaller values. This indicates a sequential spread of quantum correlations which are also observed in the bound pairs in the subradiant coupling regime \cite{Jen2021_bound}. Here by contrast, the quantum correlations maintain finite values even away from the subradiant coupling regime ($\xi\sim\pi$) and are thus purely assisted by disorders.  

\begin{figure}[b]
\centering
\includegraphics[width=8.5cm,height=4.5cm]{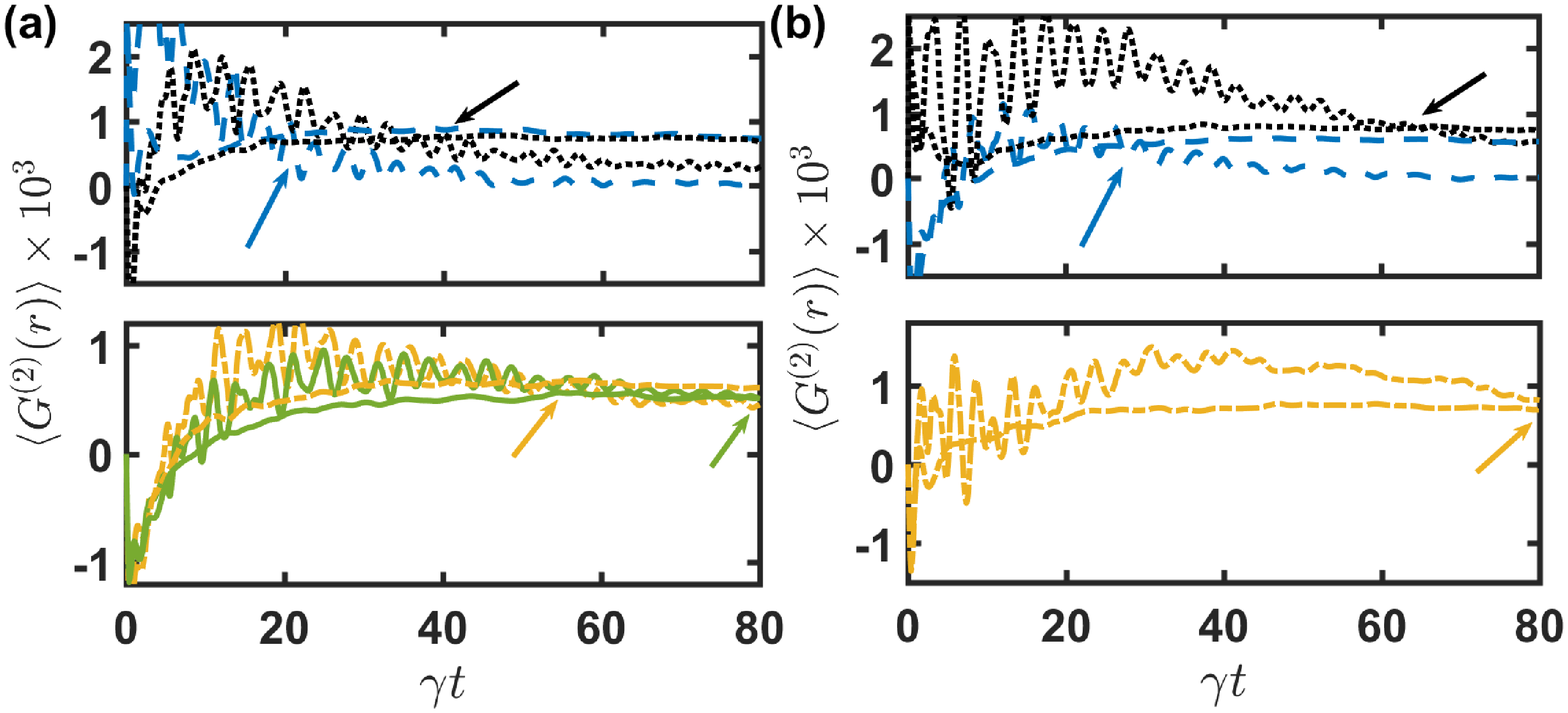}
\caption{Crossing times in $\langle G^{(2)}(r)\rangle$. Various approximate crossing times of $\langle G^{(2)}(r)\rangle$ between the cases under disorders (leveled at long time) and without (decaying within time of interest) are located by the arrows for initial double atomic excitations with separated lattice sites of (a) one ($r=2$) and (b) two ($r=3$), respectively. Various line styles are the same as in Fig. \ref{fig4}(a). A longer-range correlation with a larger $r$ takes a longer crossing time to sustain over the ones without disorders.}\label{fig6}
\end{figure}

Next we investigate in detail the crossing times of various $\langle G^{(2)}(r)\rangle$. In Fig. \ref{fig6}, we take an example of initial double excitations separated by one and two lattice sites, respectively. As expected, $\langle G^{(2)}(r=2)\rangle$ and $\langle G^{(2)}(r=3)\rangle$ present the maximums at long time comparing other quantum correlations at different $r$. When investigating further the crossing times after which the correlations without disorders decay and vanish, these crossing times are delayed for longer-range quantum correlations. This shows how correlations scramble through the atomic chain with a finite time, and a complete build-up of mutual quantum correlations between each atoms would take an ultimate time span determined by two farthest atoms in the chain.   

\begin{figure}[t]
\centering
\includegraphics[width=8.5cm,height=4.5cm]{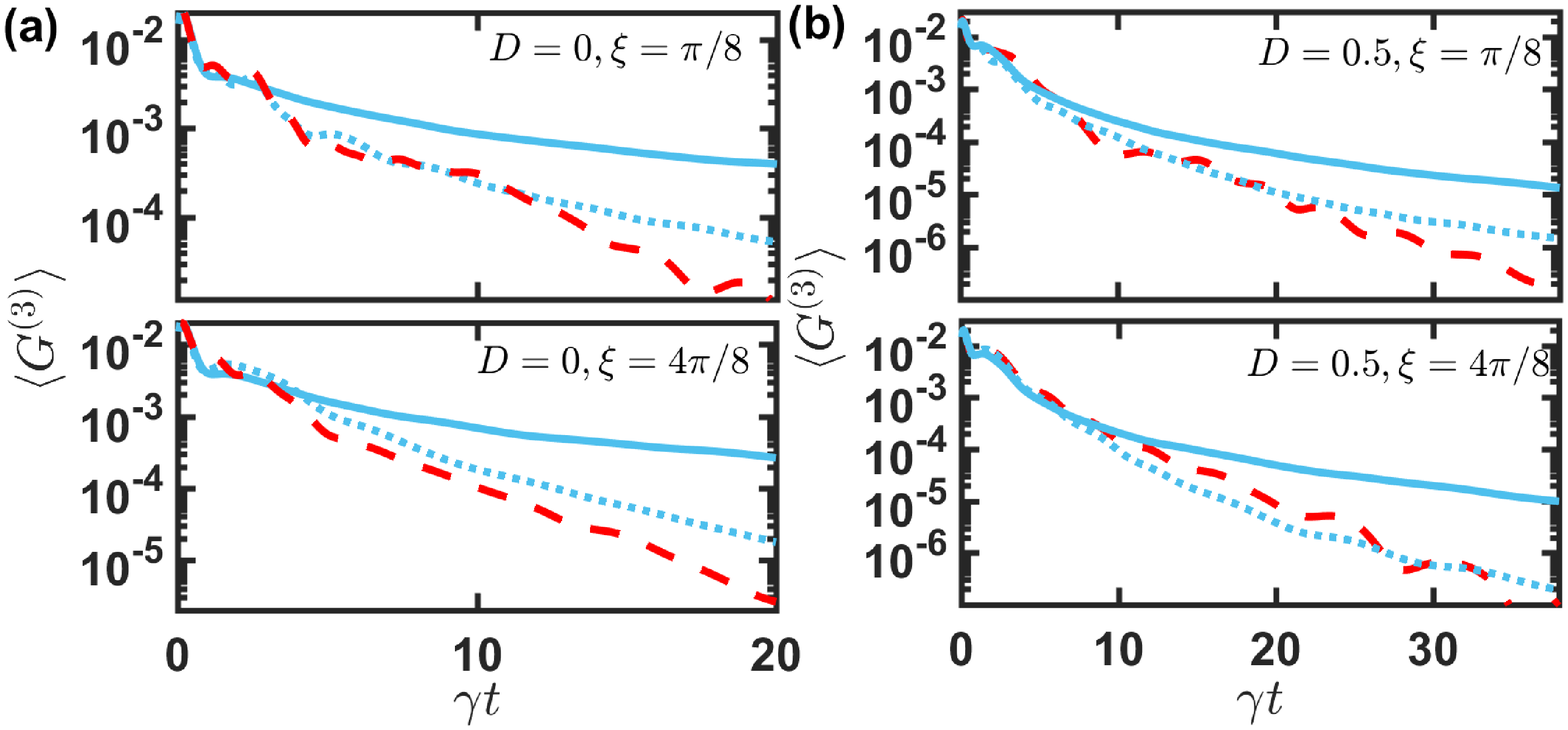}
\caption{Average third-order quantum correlation $\langle G^{(3)}\rangle$. A comparison of crossing times for different directionality $D$ and interaction strength $\xi$ is demonstrated, where various line styles correspond to the cases in Fig. \ref{fig3}, which are $W/\pi$ $=$ $0$ (dash line), $0.1$ (dotted line), and $0.8$ (solid line).}\label{fig7}
\end{figure}

As for three atomic excitations, in Fig. \ref{fig7} we compare $\langle G^{(3)}\rangle$ for various $D$ and dipole-dipole interaction strengths determined by $\xi$. Similar to $\langle G^{(2)}(r)\rangle$, they show the same behavior as the second-order correlations, which are suppressed initially in short time and surpass the one without disorder at a later stage. For a finite $D$ in Fig. \ref{fig7}(b), a larger $\xi$ tends to delay the crossing time for both strong and moderate disorder strengths. A particular slow slope emerges when $\langle G^{(3)}\rangle$ passes through the crossing point of time. On the other hand for the case of reciprocal coupling in Fig. \ref{fig7}(a), a larger $\xi$ does not significantly shift the crossing time, but it stays along with a similar slope until $\gamma t\sim 10$, close to the crossing time when $\xi$ is smaller. Nonetheless, $\langle G^{(3)}\rangle$ decreases for a larger $\xi$ under a moderate disorder strength of $W/\pi=0.1$, which coincides with the results in Sec. III that the interactions drive the system toward a delocalized side for low disorder strengths. Under a stronger disorder as shown in Fig. \ref{fig7}, $\langle G^{(3)}\rangle$ as a bulk property does not modify significantly as $\xi$ increases. 

\section{Discussion and conclusion}

To fully understand a delocalization to localization transition in open quantum many-body systems is challenging. The challenge is at least twofold. One is the thermodynamic limit in numerical simulations, which can not be easily achieved when a system is under strong atom-atom interactions. For a finite-size system, a crossover of non-Hermitian many-body localization transition \cite{Hamazaki2019} can only be identified, where a crossover behavior also shows up in a Anderson-like localization transition in terms of level statistics \cite{Jen2021_crossover}. The other restriction is the simulation or observation time in theoretical modelings or experiments, respectively, since localization of quantum particles may take a longer time than the one our methodologies can allow. In particular, it would be more difficult to unveil the localization mechanism when the system is under dissipation, where the quantum particles may decohere and relax before the localization sets in.    

Therefore, we take an alternative approach with an aid of quantum correlations in time to probe the system dynamics in localization of particles. The all-order quantum correlations can ultimately reveal the complete information of system's evolutions from quenched atomic excitations. Here we focus on the atom-waveguide interface, where atoms can align near the optical fiber \cite{Corzo2019} and form a strongly interacting system through the confined guided modes \cite{Arcari2014, Tiecke2014, Yala2014}. This interface under chiral couplings \cite{Lodahl2017} can present intriguing properties of light, for example, modified collective radiations or nonreciprocal absorptions, and as well offer many new opportunities in studying nonequilibrium many-body dynamics and localization of spin excitations in a disordered media. Finally, a potential platform using two parallel nanofibers \cite{Kien2020} promises a quantum optical setup of two-dimensional waveguide quantum electrodynamics that can host flat bands and tailor the properties of light correlations \cite{Marques2021}. 
 
In conclusion, we have demonstrated the time-evolving quantum correlations of localized atomic excitations in a disordered atomic chain. From a localization side, a crossing of time evolutions for the average quantum correlations presents a distinct separation of two regimes indicating the onset of the excitations localization. We have shown a disorder-assisted build-up of quantum correlations that can sustain for long time owing to the absence of excitations diffusion. The quantum correlations are highly correlated to the initialization of the system, which can serve as complimentary measures to potentially probe the few-body and many-body localization transitions. 

\section*{ACKNOWLEDGMENTS}

We acknowledge support from the Ministry of Science and Technology (MOST), Taiwan, under the Grant No. MOST-109-2112-M-001-035-MY3. We are also grateful for support from TG 1.2 and TG 3.2 of NCTS and inspiring discussions with Yi-Cheng Wang and Jhih-Shih You. 


\end{document}